\documentclass[aps,article,nofootinbib,twocolumn]{revtex4}

\usepackage{epsfig}
\usepackage{graphicx}
\usepackage{graphics}

\usepackage{amsfonts}  
\usepackage{amsmath, amssymb, stmaryrd}     

\usepackage{psfrag}  

\usepackage{pict2e}
\usepackage{mdwlist}
\def\be{\begin{equation}}
\def\ee{\end{equation}}
\def\bi{\begin{itemize}}
\def\ei{\end{itemize}}

\usepackage{multirow}  
\usepackage{slashbox}

\begin{document}
\title{Domain walls coupled to matter: the symmetron example}
\author{Claudio Llinares}
\affiliation{Institute of Theoretical Astrophysics, University of Oslo, 0315 Oslo, Norway}
\author{Levon Pogosian}
\affiliation{Department of Physics, Simon Fraser University, Burnaby, BC, V5A 1S6, Canada}

\begin{abstract}

We study properties of domain walls in the symmetron model, in which the scalar gravitational degree of freedom decouples from matter in regions of high density, and exhibits a spontaneously broken $Z_2$ symmetry at low densities. The non-minimal coupling of the scalar to matter leads to a host of interesting properties of the domain walls that are not present in minimally coupled theories. We estimate the cosmological energy fraction in domain walls and find that this leads to an upper bound on the redshift of the symmetry breaking. We also show that a spherical symmetron wall can remain stable if it is ``pinned'' on matter halos and derive a criterion for the stability. In addition, we present results of numerical simulations of representative interactions between domain walls and matter over-densities.

\end{abstract}

\maketitle

\section{Introduction}

Presence of topological defects, or soliton configurations of any type, is an important consideration in building models of particle physics and cosmology. Their presence in the theory provides another, often very effective, handle on constraining the allowed parameter space. The evolution and interactions of defects, such as monopoles, cosmic strings and domain walls, have been subjects of many studies \cite{2000csot.book.....V}, revealing a multitude of fascinating properties owing to their highly non-linear nature.  The formation and scaling of a network of defects strongly depends on how they interact among themselves and with the environment. This, in turn, affects the type and the strength of restrictions that observations (or the lack thereof) can impose on the underlying model. 

In the context of cosmological domain walls, there is a large amount of work on understanding their dynamics and evolution based on both analytical \citep{1996PhRvL..77.4495H, 2003PhRvD..68d3510H, 1984PhRvD..30.2036V} as well as numerical techniques \citep{1989ApJ...347..590P, 1990PhRvD..41.1013K, 1996PhRvD..53.4237C, 1997PhRvD..55.5129L, 2003PhRvD..68j3506G, 2005PhRvD..72h3506A, 2011PhRvD..84j3523L, 2013PhLB..718..740L}.  Their interactions with each other and with monopoles \cite{Pogosian:1999zi,Pogosian:2000xv,Pogosian:2001pq,Pogosian:2002ua,Antunes:2003be}, as well as their gravitational effects \citep{1981PhRvD..23..852V, 1989PhRvD..39.3571W, 1985PhLB..162..287T, Ipser:1983db}, including emission of gravitational waves \citep{2014JCAP...02..031H}, are well studied. All of these studies have assumed no direct coupling between the order parameter, typically a scalar field, and matter. The reason for this assumption is simple -- any non-minimal coupling with matter would amount to presence of new gravitational degrees of freedom mediating fifth forces, which are strongly constrained by Solar System test of gravity. However, the discovery of Cosmic Acceleration, coupled with the Old Cosmological Constant Problem and the unexplained nature of Dark Matter, caused a surge of interest in various modifications of gravity \cite{Silvestri:2009hh,2012PhR...513....1C,Joyce:2014kja}. Since practically all of the proposed models contain extra gravitational interactions, they must also include a screening mechanism that effectively switches off the fifth force on Earth and the Solar System, thus restoring General Relativity in regions where it has been tested the best.  In this paper, we consider an example of a such a theory, the symmetron model, in which interactions with matter lead to some new interesting properties of domain walls (DWs).  We note that qualitatively similar phenomena are well-known in the context of ferromagnets, where domain walls interact with impurities (see, {\it e.~g.} \cite{jiles_hysteresis} and references therein).

The symmetron model is a scalar-tensor theory of gravity, proposed in \cite{2010PhRvL.104w1301H}, in which the scalar field decouples from matter when the matter density is high. Below a certain critical density, a $Z_2$ symmetry is spontaneously broken giving a non-zero vacuum expectation value (VEV) to the scalar field.  The cosmology of this model at the background and linear perturbation level has been studied in \cite{2011PhRvD..84j3521H,2011PhRvD..84l3524B}.  In the non-linear case, there are several results coming from quasi-static non-linear N-body cosmological simulations \cite{2012ApJ...748...61D, 2012JCAP...10..002B, 2014A&A...562A..78L}.  The model was shown to be capable of leaving its fingerprints in observables such as gravitational redshift \cite{2014A&A...562A...9G} and the shape of galaxy clusters \cite{2013PhRvL.110o1104L}.  As in any theory with a spontaneously broken discrete symmetry, the symmetron model contains DW solutions connecting regions that happen to pick different VEV. Such domain walls were recently observed to form in {\em non-}quasi-static N-body simulations of structure formation \cite{2013PhRvL.110p1101L, 2014PhRvD..89h4023L}.

In what follows, we study in detail properties of symmetron domain walls, starting from evaluating their thickness and the surface energy density. We then proceed to estimate the fraction of energy in such domain walls for viable symmetron models. This fraction, and other properties, depend on the redshift at which the $Z_2$ symmetry was broken. In particular, the fraction of energy density in DWs becomes large for large values of $z_{SSB}$.  This would leave an imprint on the evolution of the background, as well as the growth of structures. Hence, taking domain wall formation into account provides an upper bound on the redshift of the symmetry breaking. We derive a stability condition for a spherical wall pinned on matter halos. Then, using numerical simulations, we study a few representative interactions between domain walls and congregations of matter, demonstrating how a wall becomes stabilized by attaching itself to matter filaments. We note that, while our paper was in preparation, another paper studying similar configurations has appeared \cite{2014arXiv1409.6570P}.  Finally, we refer the reader seeking a concise review of properties of conventional domain walls to the books by \citet{2000csot.book.....V} and \citet{2006kdwi.book.....V}. 
\section{The symmetron model}

The action of the symmetron model is given by \cite{2010PhRvL.104w1301H}
\be
S = \int \sqrt{-g} \left[ R - \frac{1}{2}\nabla^a\phi \nabla_a \phi - V(\phi)\right] d^4x + S_M(\tilde{g}_{ab}, \psi) \ ,
\label{symm-action}
\ee
where the Einstein $g_{ab}$ and the Jordan $\tilde{g}_{ab}$ frame metrics are related via
\be
\tilde{g}_{ab} = A^2(\phi) g_{ab},
\ee
and the potential and the conformal factor have the following forms:
\begin{align}
\label{potential}
V(\phi) &= -\frac{1}{2}\mu^2\phi^2 + \frac{1}{4}\lambda\phi^4 + V_0 \\
\label{conformal_factor}
A(\phi) &= 1 + \frac{1}{2}\left(\frac{\phi}{M}\right)^2 \ ,
\end{align}
where $\mu$ and $M$ are mass scales, $\lambda$ is a dimensionless constant and $V_0$ is tuned to match the observed cosmological constant. The equation of motion of the scalar field is
\be
\nabla^a\nabla_a\phi = V_{,\phi} - A^3(\phi) A_{,\phi} \tilde{T}, 
\label{eq_motion_phi}
\ee
where 
\be
\tilde{T}_{ab} = -2\frac{1}{\sqrt{-\tilde{g}}} \frac{\delta L_M}{\delta\tilde{g}^{ab}}
\ee
is the Jordan frame energy momentum tensor. We adopt a perturbed Friedmann-Robertson-Walker metric which in the Einstein frame is given by
\be
ds^2 = -(1+2\Phi) dt^2 + a^2(1-2\Phi)(dx^2+dy^2+dz^2) .
\ee
In the quasi-static limit, the scalar field obeys the following equation of motion:
\be
\nabla^2\phi = \left(\frac{\rho}{M^2}-\mu^2 \right)\phi + \lambda \phi^3 = \frac{d}{d\phi}V_{eff}(\phi), 
\ee
where $\rho$ is the Jordan frame matter density and the effective potential is given by
\be
V(\phi)_{eff} = \frac{1}{2}\left(\frac{\rho}{M^2} - \mu^2\right)\phi^2 + \frac{1}{4}\lambda\phi^4 + V_0.
\label{def_effective_potential}
\ee
From the form of $V_{eff}(\phi)$, one can see that the expectation value of the scalar field vanishes at high matter densities, setting the conformal factor $A$ to unity and decoupling the scalar from the matter. Thus, the symmetron model is an example of a theory with a screened fifth force with the benefit of having a renormalizable potential.

It is convenient to work with a dimensionless scalar field $\chi \equiv \phi/\phi_0$, where $\phi_0$ is the expectation value at zero matter density:
\be
\phi_0 = \frac{\mu}{\sqrt{\lambda}}.
\ee
Also, it helps to relate the 3 free parameters of the model, $(\mu, \lambda, M)$, to the Compton wavelength of the scalar field at $\rho=0$,
\be
\lambda_0 = \frac{1}{\sqrt{2}\mu} \ , 
\label{def_lambda0}
\ee
a dimensionless coupling constant,
\be
\beta = \frac{\phi_0 M_{pl}}{M^2} \ ,
\label{def_beta}
\ee
and the scale factor at time of the symmetry breaking,
\be
a_{SSB}^3 = \frac{\rho_0}{\rho_{SSB}} = \frac{\rho_0}{\mu^2 M^2} \ , 
\label{def_assb}
\ee
where $\rho_0$ is the background density today, at $a=1$.  Throughout the paper we will use both  $a_{SSB}$ or its associated redshift $z_{SSB}$.  Eqs.~(\ref{def_lambda0}), (\ref{def_beta}) and (\ref{def_assb}) imply the following reverse relations between the original and the new parameters:
\begin{align}
\label{inverse_m}
M^2 & = 2\lambda_0^2 \rho_{SSB} \\
\label{inverse_mu}
\mu^2 & = \frac{1}{2\lambda_0^2}\\
\label{inverse_lambda}
\lambda & = \frac{M_p^2}{8\lambda_0^6 \beta^2 \rho_{SSB}^2} = \frac{1}{2\lambda_0^2\phi_0^2}. 
\end{align}
Combining the equations above leads to the following useful expression for $\phi_0$:
\be
\phi_0 = \frac{2\lambda_0^2 \beta}{M_{pl}} \rho_{SSB} = 6 \frac{\lambda_0^2 \beta}{a_{SSB}^3} H_0^2 \Omega_0 M_{pl}.
\label{phi0_of_params}
\ee

The equation for the dimensionless scalar field $\chi$ is
\be
\nabla^2\chi = \frac{a^2}{2\lambda_0^2}\left[\left(\frac{\rho}{\rho_{SSB}} - 1\right)\chi + \chi^3 \right] \ 
\label{eq_motion_chi}
\ee
with the following solutions in the homogeneous matter density limit:
\be
\label{solution_constant_rho}
\chi = \begin{cases}
0 & \text{if $\rho$ $>$ $\rho_{SSB}$ (screened)},\\
\left(1-\frac{\rho}{\rho_{SSB}} \right)^{1/2} & \text{if $\rho$ $<$ $\rho_{SSB}$ (un-screened)}.
\end{cases}
\ee

\subsection{Allowed parameter space}
\label{allowed_parameters}

Cassini measurements of the ratio between perturbations of the time and space components of the metric imply the following bounds on parameters of the symmetron model \cite{2010PhRvL.104w1301H}:
\begin{align}
M & \lesssim 10^{-3} M_p, \\
\beta & \sim 1.
\label{allowed_models_M}
\end{align}
This translates into a constraint on a combination of $\lambda_0$ and $a_{SSB}$:
\be
\frac{\lambda_0}{a_{SSB}^{3/2}} \lesssim 10^{-3} H_0^{-1} \ .
\label{allowed_models}
\ee
Thus, assuming we restrict to models with coupling strengths of order unity, one can have viable symmetron models with larger scalar interaction range if the symmetry is broken at low redshifts, or the smaller range if the $z_{SSB}$ is high. As our fiducial model, we take the one defined by parameters $(\lambda_0, z_{SSB}) = (1 \mathrm{Mpc}/h, 1)$, which was discussed in \cite{2010PhRvL.104w1301H} and simulated in \cite{2012ApJ...748...61D, 2014A&A...562A..78L, 2014PhRvD..89h4023L}.  The fifth force associated with this model has a range comparable to the size of a typical dark matter halo.  We will also be interested in using models that have a smaller range (and thus a larger value of $z_{SSB}$), since, as we will show later, they lead to observationally interesting values of the energy density fraction in the domain walls.

It follows from Eqs.~(\ref{allowed_models_M}) and (\ref{allowed_models}) that the allowed departure of the conformal factor from unity is very small, even in the vacuum limit:
\be
A(\phi_0) = 1+ {\cal O}(10^{-6}) \ .
\label{asim1}
\ee

\section{Properties of the symmetron domain walls}

\subsection{The DW solution for homogeneous matter densities}

The profile of a planar domain wall can be calculated analytically for homogeneous matter densities. Assuming $\rho < \rho_{SSB}$ and solving Eq.~(\ref{eq_motion_chi}) with the boundary conditions (see Eq.~(\ref{solution_constant_rho}))
\begin{equation}
\chi_{\pm \infty}  = \pm \left( 1 - \frac{\rho}{\rho_{SSB}} \right)^{1/2} 
\label{bc}
\end{equation}
we obtain a solution
\be
\chi_{dw}(x,a) = \sqrt{1 - \frac{a_{SSB}^3}{a^3}} \tanh\left[\frac{ax}{2\lambda_0} \sqrt{1 - \frac{a_{SSB}^3}{a^3}} \right] \ ,
\label{dw-solution}
\ee
where we used the fact that $\rho \propto a^{-3}$. Here we find the first (obvious) difference from conventional domain walls: there is a dependence of the solution on the matter density and, hence, the scale factor. Note that the expression is only valid for $a>a_{SSB}$. At $a<a_{SSB}$, the scalar field has a zero VEV and there are no domain walls.

\subsection{Thickness}

\begin{figure}[tb]
    \includegraphics[width=0.45\textwidth]{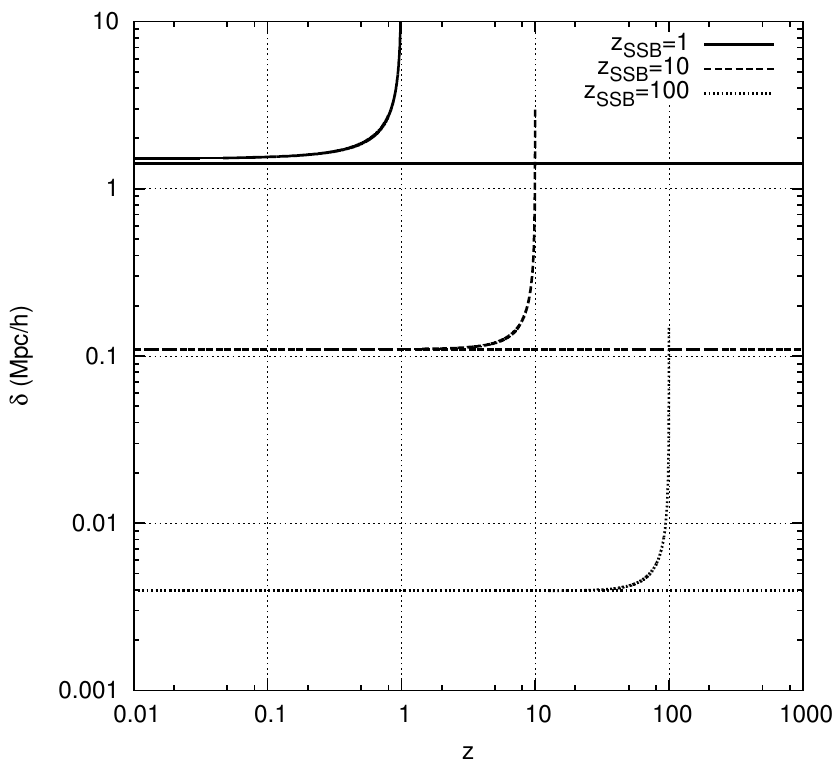}
    \caption{Physical thickness of symmetron domain walls as a function of redshift for three different sets of model parameters. The curves that increase with redshift correspond to the walls in the symmetron model, while the horizontal straight lines are the values for ``conventional'' domain walls, when $\rho=0$.}
    \label{fig:thickness_of_z}
\end{figure}

\citet{2000csot.book.....V} define the width of the wall as the value of the coordinate at which the argument of the $\tanh$ function is equal to $1/\sqrt{2}$.  Using the same definition applied to the physical coordinate $ax$, we obtain 
\be
\delta = \sqrt{2}\lambda_0 \left( 1 - \frac{a_{SSB}^3}{a^3} \right)^{-1/2}.
\ee
Note that this is the physical thickness and the comoving thickness would have an additional $1/a$ factor. The thickness has a dependence on time that does not exist for conventional walls. Fig.~\ref{fig:thickness_of_z} shows representative values of $\delta$ at various redshifts $z$ for three different models characterized by the redshift of symmetry breaking $z_{SSB}$.  The parameter $\lambda_0$ was determined from Eq.~(\ref{allowed_models}). The width of the walls diverges at the instant they are born, $z=z_{SSB}$.  After that, the width decreases rapidly until it reaches the vacuum value limit.  Note that this plot was made assuming that the density is equal to the mean density.  In the following sections we will see that domain walls actually prefer to be pinned on matter over-densities and, therefore, their widths will, in general, be larger than the width of a wall in the vacuum.

\subsection{Energy content}

\subsubsection{Surface energy density of a planar wall}

Because of the coupling with matter, the energy trapped in a DW depends not only on the configuration of $\phi$ but also on $\rho$. We define the energy density due to a DW as the difference between $-\tilde{T}^0_0$ of a universe with a DW minus $-\tilde{T}^0_0$ of a universe without a DW. To start, we calculate the surface energy density of a planar wall in the Jordan frame, given by
\be
\sigma = \int_{-\infty}^{\infty} a \ dx \ [-\tilde{T}^0_0(\mathrm{DW}) + \tilde{T}^0_0(\mathrm{background})] \ .
\label{sigma}
\ee
The energy momentum tensor used in this definition can be obtained by varying the scalar field and the matter parts of the action (\ref{symm-action}) with respect to the Jordan frame metric $\tilde{g}_{ab}$ and is given by
\begin{align}
\nonumber
\tilde{T}_{ab} &= A^{-2}(\phi) \left[ \nabla_a\phi \nabla_b\phi - \frac{1}{2}g^{cd}\nabla_c\phi\nabla_d\phi g_{ab}
- V(\phi)g_{ab} \right] \\ & + \tilde{T}^{M}_{ab} \ .
\end{align}
The energy density component is
\be
-\tilde{T}^0_0  = A^{-4}(\phi)\left[\frac{1}{2a^2}|\nabla\phi|^2 + V(\phi)\right] + \rho \ , 
\label{t00}
\ee
where we used the fact that time derivatives vanish for a static configuration. Substituting (\ref{t00}) into (\ref{sigma}), and ignoring the $A^{-4}$ prefactor which is indistinguishable from unity in viable models (see (\ref{asim1})), we can write
\begin{multline}
\sigma = \phi_0^2  \int_{-\infty}^{\infty} adx \left\{ \frac{1}{2a^2} |\nabla \chi_{dw}|^2 \right.\\
\left.  + {1 \over 8 \lambda_0^2}
\left[ \chi_{dw}^4 - \chi_{\infty}^4 - 2(\chi_{dw}^2 - \chi_{\infty}^2)  \right] \right\},
\end{multline}
where $\chi_{\infty}$ and $\chi_{dw}$ are given by (\ref{bc}) and (\ref{dw-solution}), respectively. The integral can be readily evaluated, giving
\be
\sigma = \frac{2\phi_0^2}{3\lambda_0} \ \gamma(a,a_{\rm SSB})
\label{sigma-dw}
\ee
with
\be
\gamma(a,a_{\rm SSB}) \equiv \sqrt{1-\frac{a_{SSB}^{3}}{a^{3}}} \left[ 1+ {a^3_{SSB} \over 2a^3}\right] \ .
\label{gamma}
\ee
Eq.~(\ref{sigma-dw}) reduces to the standard expression for the DW surface energy, $\sigma=2\phi_0^2/(3\lambda_0)=2\sqrt{2}\mu^3/(3\lambda)$ \cite{2006kdwi.book.....V}, in the $a_{SSB} \rightarrow 0$ or $\rho \rightarrow 0$ limits. For the symmetron DW, the surface energy evolves with time because of its dependence on the background density $\rho$.

\subsubsection{Energy density of a set of walls}

\begin{figure}[tbh]
\includegraphics[width=0.45\textwidth]{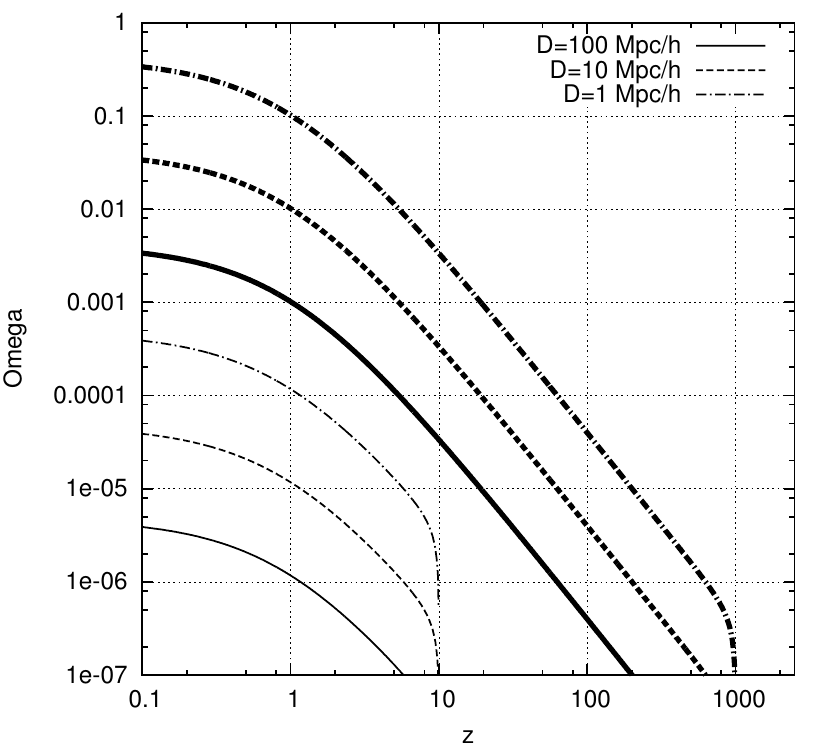}
\caption{The ratio, $\Omega_{\rm DW}$, of the DW energy density and that of matter for a set of parallel domain walls separated by a comoving distance $D$. The thin and thick lines correspond to $z_{SSB}=10$ and $z_{SSB}=1000$, respectively.  The continuous, dashed and dotted-dashed lines corresponds to values of $D$ of 100, 10 and 1 Mpc/h, respectively.}
\label{fig:omega_of_z}
\end{figure}

Given the surface energy for one wall, we can estimate the mean cosmological energy density of a set of walls.  For this, at first, we assume a simple model with parallel walls separated by some distance that grows in proportion with the scale factor.  We define this physical distance as $a D$, where $D$ is a constant.  Then, the mean DW energy density is
\be
\rho_{\mathrm{DW}} = \frac{\sigma}{a D}.
\label{rho_as_f_of_D}
\ee
Note that the scaling $1/a$ is consistent with an equation of state $w=-2/3$. One way to quantify the relative contribution of DW to the energy budget is to define $\Omega_{\rm DW}$ as the ratio of the domain wall energy density to that of matter. We write
\be
\Omega_{\rm DW} \equiv {\rho_{\mathrm{DW}} \over \rho} = 8\beta^2 \Omega_M {H_0^2\lambda_0^3 \over Da_{SSB}^6} a^2 \gamma(a,a_{\rm SSB}) \ ,
\ee
where $\Omega_M$ is the matter fraction today. Taking into account the constraint (\ref{allowed_models}) on the allowed parameter space for symmetron models with $\beta \sim 1$ gives
\be
\Omega_{\rm DW} \lesssim 10^{-8}  (H_0D)^{-1} a^{-3/2}_{SSB} \Omega_M \ a^2 \gamma(a,a_{\rm SSB}) 
\label{omega-dw}
\ee
Fig.~\ref{fig:omega_of_z} shows the time evolution of $\Omega_{\rm DW}$ for several models at the boundary of the allowed symmetron space for different values of $z_{SSB}$ and $D$.

So far, we treated the distance $D$ between the walls as a free parameter. In the next subsection, we estimate it based on the process by which domain walls form in the symmetron model.

\subsubsection{Separation between the walls from semi-analytic arguments}

In the symmetron model, DWs form when the density in a given void drops below the density of symmetry breaking for the first time. At that moment, there is a wave moving away from that point which sets up the values of the field away from zero and fixes the sign.  A wall is formed when this wave finds another wave coming from a different void that also reached the density of symmetry breaking. Thus, the size of the domains is related to the distance between regions in which symmetry is broken for the first time. Note that this process of wall formation is different from the process associated with standard domain walls in which only the background values are taken into account and, thus, a typical separation between the walls is always of the order of the horizon. 

Since the wall formation depends on the density distribution at a given time, we will use realizations of a given matter power spectrum in a box.  We used the open source code CAMB \cite{Lewis:1999bs} to produce a $\Lambda$CDM power spectrum. It is then used to generate random Gaussian realizations in Fourier space that are subsequently converted into realizations of the density field \cite{llinares_thesis}.  At subsequent times, we assume a linear evolution, which means 
\be
\delta(t,x) = \mathcal{D}(t)\delta(z=0,x),
\ee
where $\mathcal{D}(t)$ is normalized to unity at redshift zero.  For the purpose of this crude calculation, which only aims at an order of magnitude estimate of the inter-wall separation $D$, we can safely assume an Einstein-de Sitter universe and $\mathcal{D}(a)=a$.

To estimate $D$, we need to estimate the distance between a point of minimum density  $x_{min}$ and the rest of the points for which the symmetry can be broken before the wave coming from $x_{min}$ arrives.  In order to do this, we need to find the local time for symmetry breaking ($t_{LSSB}$), which is defined as the solution of
\be
\rho(t,x) = \rho_{SSB}.
\ee
Switching to the density contrast $\delta \equiv \rho(x,t)/\rho$, we get an equation for the expansion factor for local symmetry breaking ($a_{LSSB}$):
\be
a_{LSSB}^3 - \delta(z=0,x)a_{SSB}^3 a_{LSSB} - a_{SSB}^3=0, 
\ee
which has a unique analytic solution if $\delta(z=0,x)<0$ (i.e. in voids).  Finally, we need to compare this time with the time needed for a wave to arrive from $x_{min}$.  This can be found from the definition of the null geodesics
\be
\frac{dt}{dx} = \frac{a}{c} \ ,
\ee
with the solution given by
\be
\int_{t_{min}}^t t^{-2/3}dt = \int_0^x \frac{1}{c}\left(\frac{3}{2}H_0\right)^{2/3} dx, 
\ee
where we defined $t_{min}$ as the time at which the symmetry in broken at the point $x_{min}$.  The solution for the null geodesics is 
\be
t_{ng} = \left[ \frac{1}{3c}\left( \frac{3}{2}H_0\right)^{2/3}x + t_{min}^{1/3} \right]^3.
\ee
which allows us to write a criteria for wall formation: a given point will be a potential seed for a wall if it fulfills 
\be
t_{ng}>t_{LSSB}.
\label{arrival_wave}
\ee

\begin{figure}
  \includegraphics[width=0.45\textwidth]{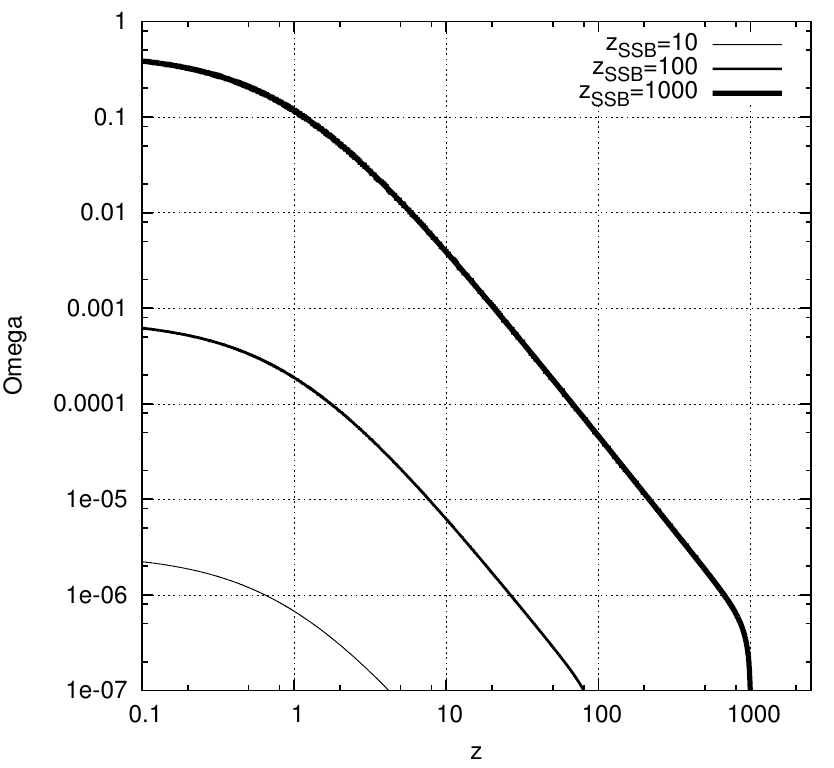}
  \caption{The ratio, $\Omega_{\rm DW}$, of the DW energy density and that of matter given by Eq.~(\ref{omega-dw}) using the distance between walls given by Eq.~(\ref{arrival_wave}).  Different thickness of the lines correspond to different values of $z_{SSB}$.}
\label{fig:omega_of_z_fixed_d}
\end{figure}

\begin{table}[tb]
  \begin{center}
  \begin{tabular}{cc}
    $z_{SSB}$ & D (Mpc/h) \\
    \hline 
       1 & 800 \\
      10 & 200 \\
     100 & 20 \\
    1000 & 1
  \end{tabular}
 \caption{Estimated distance $D$ between domain walls obtained using Eq.~(\ref{arrival_wave}) for several models defined by their $z_{SSB}$ (note that this is the only model parameter that enters in the calculation).}
  \label{tab:values_of_d}
  \end{center}
\end{table}

We applied this criterion to determine the distance between walls of several models defined by different values of $z_{SSB}$.  To this end, we used several realizations of the over-density field at different resolutions and box sizes to make sure that the estimates are independent of these quantities. Representative values are shown in Table~\ref{tab:values_of_d}.

Non-quasi-static simulations of structure formation in the symmetron model were presented in \cite{2014PhRvD..89h4023L} for a model with $z_{SSB}=1$ in box of size 128 Mpc/h.  They found one domain wall with a size of the order of half the box size, which is smaller than our best estimate for the size of the walls for this particular model.  Results of this section imply that in order to accurately model domain wall formation the simulation should be run in a much larger box.  On the other hand,  the solar system constraints limit the range of the scalar field to less than 1 Mpc/h for this value of $z_{SSB}$.  This means that a realistic simulation should not only have a large box, but also a spacial resolution of at least 1 Mpc/h, which represents a serious computational challenge.

Finally, in Fig.~\ref{fig:omega_of_z_fixed_d}, we show the evolution of $\Omega_{\rm DW}$ for several values of $z_{SSB}$ and under the same assumptions as in Fig.~\ref{fig:omega_of_z}, but with the average inter-wall distance $D$ estimated using the semi-analytical arguments of this subsection. $\Omega_{\rm DW}$ becomes large for large values of $z_{SSB}$, which would affect the evolution of the background, as well as the linear and non-linear growth of structures. Thus, taking domain wall formation into account provides an upper bound to the model parameter $z_{SSB}$, for which the present literature does not provide any constraint. 

\section{Domain walls pinned to matter over-densities}

Unlike the conventional domain walls,  the symmetron walls prefer to stay in places where the density is high. This can be seen from simple energy arguments, and happens for two reasons:
\begin{enumerate}
\item The energy of the walls decreases with increasing matter density (see the dependence on $\rho$ in Eq~(\ref{sigma-dw})). Thus, they tend to move away from low density regions and stay close to over-densities.
\item In places where the density grows beyond the necessary value to restore the symmetry, the value $\phi=0$ becomes a minimum of the effective potential and thus a stable point.  In these places, the wall does not sit anymore in a maximum of the potential, but a minimum, becoming more stable.
\end{enumerate}
Cosmological simulations presented in \cite{2013PhRvL.110p1101L} and \cite{2014PhRvD..89h4023L} confirm this reasoning and show that realistic walls do follow dark matter structures (see, for instance, Fig.~2 in \cite{2013PhRvL.110p1101L} and Fig.~5 in \cite{2014PhRvD..89h4023L}). These simulations also show that, while domain walls tend to follow dark matter halos and stay attached to them, they also tend to collapse as time passes.  The aim of this section is to study conditions of stability for the domain walls coupled to matter. 

\subsection{Stability of spherical domain walls interacting with spherical shells of matter}

Let us start by reviewing the simple case of a spherical wall and homogeneous matter. The energy contained in such a wall of radius $R$ is given by
\be
E_{\mathrm{sphere}} = 4\pi \sigma R^2.
\label{energy_sphere}
\ee
The minimum energy principle says that the DW will evolve towards the minimization of its surface.  In the case of a sphere, this will imply that the wall will become unstable and eventually collapse.

Existence of over-densities can make spherical walls stable. To derive the condition for stability, let us assume a spherical wall located in a spherical shell that has higher density than its environment (a very primitive model for a set of dark matter halos and filaments surrounding a void).  After making a radial perturbation towards smaller radius ($R \rightarrow R-dR$), we will have the wall located in a place with smaller density (i.e. we have a change in the density $\rho \rightarrow \rho-d\rho$).  The difference between the initial and final energies is given by
\begin{multline}
dE = E(R-dR)-E(R) = \\
 CR\sqrt{1 - \frac{\rho}{\rho_{SSB}}} \left[-2\left(1-\frac{\rho}{\rho_{SSB}}\right)dR + \frac{3Rd\rho}{2\rho_{SSB}}\right],
 \label{dE}
\end{multline}
which follows from Eq.~(\ref{sigma-dw}) and a Taylor expansion, and $C$ is an irrelevant constant. In the above, for simplicity, we took into account only the gradient energy of the wall. Including the potential energy essentially contributes a factor of two to the total energy budget, which does not change the stability condition.  There are only minor differences between the gradient and potential energy that occur when the density is close to the density at symmetry breaking, which can be neglected. Eq.~(\ref{dE}) shows that the decrease in energy due to a smaller radius can be off-set by the increase caused by the change in the density. The wall will be stable if the expression inside the squared brackets is positive.  By imposing this condition, we obtain the following stability criterion for the wall:
\be
\frac{d\rho}{dR} > \frac{4\rho_{SSB}}{3R}\left(1-\frac{\rho}{\rho_{SSB}} \right).
\label{stability_criteria}
\ee
At this stage one may be tempted to substitute $\rho$ with the mean density of the universe. This will give a time dependence of the stability condition, from which one could derive a time for collapse $a_c$.  The outcome of this calculation is $a_c=a_{SSB}$. Given that the walls are born at $a_{SSB}$, we see that the mean density approximation is not good.  Realistic dark matter concentrations are much denser than the mean density and are decoupled from the expansion.  A more detailed treatment is needed to establish $a_c$.

In a realistic situation, we will not have a spherical shell, but a set of discrete halos.  In other words, a realistic shell will have holes, which effectively decrease the amount of matter that produces the stabilization of the walls.  For this reason, a spherical wall will eventually become unstable also when the holes that exist in the shell are large enough.  We study this in more detail in the following subsections.

\subsection{The second reason why walls attach to over-densities}

\begin{figure}[!t]
  \begin{center}
    \includegraphics[width=0.5\textwidth]{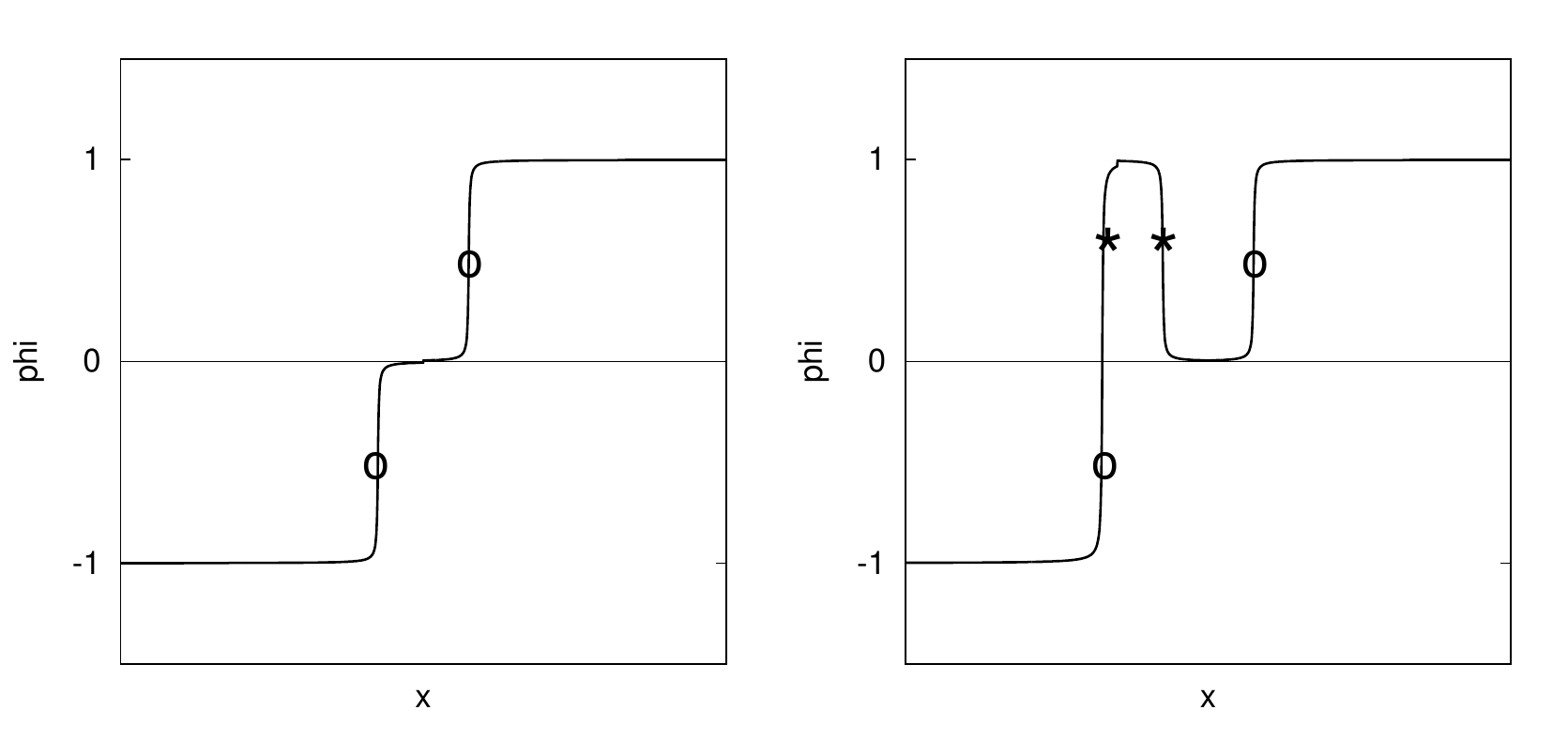}
    \caption{Scalar field corresponding to a halo and a domain wall. Left: the wall is inside the halo.  Right: the wall has moved to the left.  See text for explanation.} 
    \label{fig:energy_wall_halo}
  \end{center}
\end{figure}

In the previous subsection, we showed that the energy of the DW decreases with density, which gives a reason for domain walls to prefer high density environments.  Here we show a related reason for walls to follow halos.  The left panel of Fig.~\ref{fig:energy_wall_halo} shows a profile of the scalar field in a situation in which a wall passes through a halo.  The total gradient energy of the system (wall plus halo) is given essentially by the integral of the gradient in the regions marked with ``o''.  The right panel of the same figure shows the profile when the wall is displaced to the left.  In this case, there is an extra component of energy which is given by the integral in the regions marked with an asterisk, which did not exist before.  In order to decouple the wall from the halo one needs to give this extra energy to the system.  In other words, one needs to flip the scalar field in the halo from a negative to a positive value. The minimum energy state will correspond to the wall siting in the halo.  There are two different ways in which this extra amount of energy can be provided: from kinetic energy contained in the wall ($\dot{\phi}^2/2$) of from the energy that is released when decreasing the surface of the wall.

\subsection{Simulations in 2D}

In order to better understand the conditions for stability we run simulations using fixed density distributions.  We present results obtained using a 2D version of the code presented in \cite{2013PhRvL.110p1101L}.  The matter density is added to the grid using analytic expressions and does not evolve with time (the time scale of the evolution of the scalar field is much shorter than that of the evolution of matter, so the approximation is good).  The evolution of the scalar field is obtained by solving numerically the following 2D version of the equation of motion:
\begin{multline}
\ddot{\chi} + 3H\dot{\chi} - \frac{c^2}{a^2}\left( \frac{\partial^2\chi}{\partial x^2} + \frac{\partial^2\chi}{\partial y^2} \right) = \\
-\frac{c^2}{2\lambda_0^2} \left[ \left(\frac{a_{SSB}^3}{a^3} \eta - 1 \right) \chi + \chi^3\right],
\label{eq_motion_2D} 
\end{multline}
where $\eta$ is the local matter density normalized to the background density at the expansion factor $a$.  The 3D counterpart of these experiments will be presented in a companion paper. 

\subsubsection{Infinite DW and a set of filaments}

\begin{figure*}
\begin{minipage}[t]{0.49\textwidth}
\includegraphics[width=1.0\textwidth]{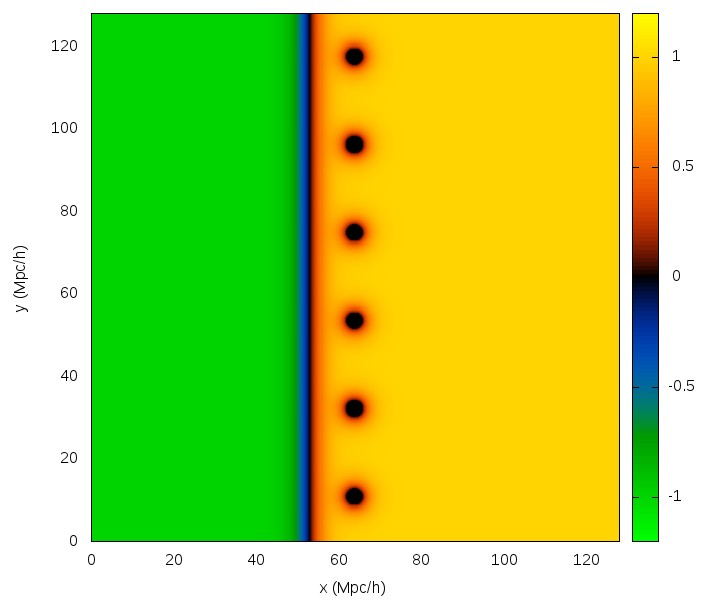}
\end{minipage}
\hfill{}
\begin{minipage}[t]{0.49\textwidth}
\includegraphics[width=1.0\textwidth]{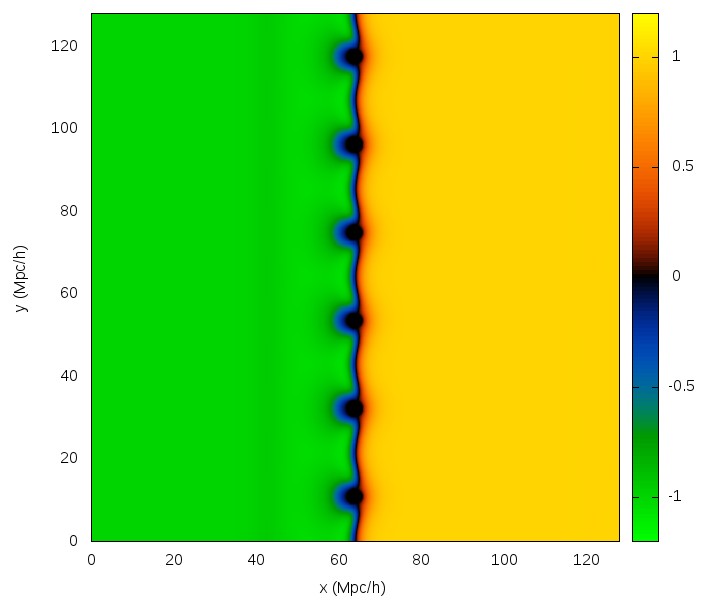}
\end{minipage}
\begin{minipage}[t]{0.49\textwidth}
\includegraphics[width=1.0\textwidth]{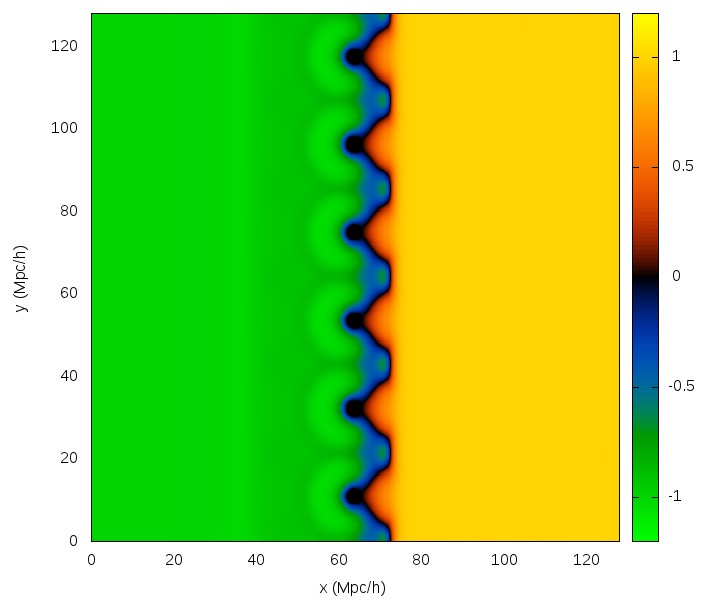}
\end{minipage}
\hfill{}
\begin{minipage}[t]{0.49\textwidth}
\includegraphics[width=1.0\textwidth]{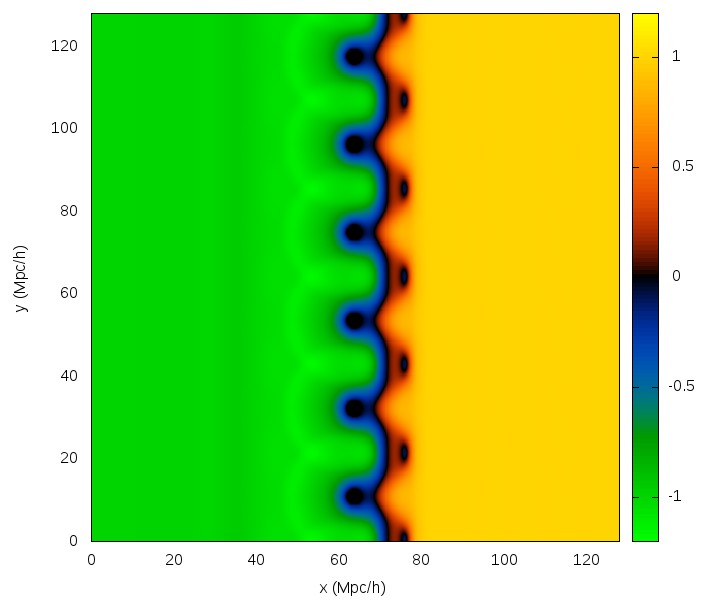}
\end{minipage}
\begin{minipage}[t]{0.49\textwidth}
\includegraphics[width=1.0\textwidth]{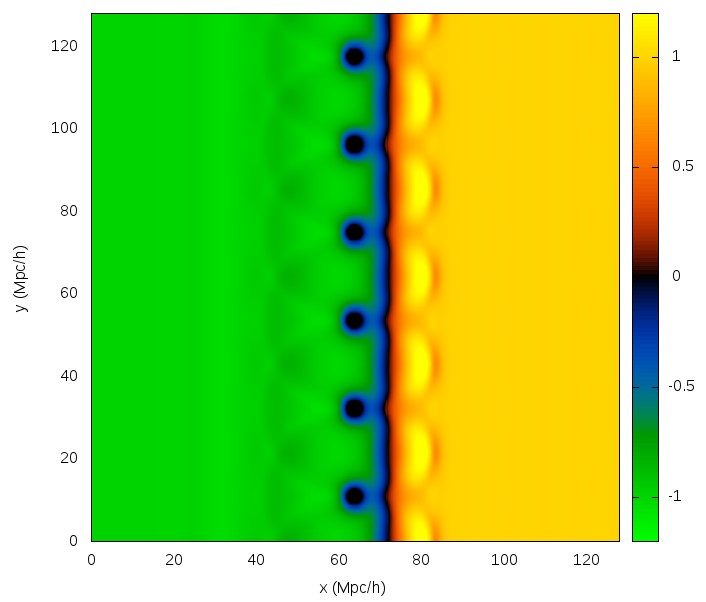}
\end{minipage}
\hfill{}
\begin{minipage}[t]{0.49\textwidth}
\includegraphics[width=1.0\textwidth]{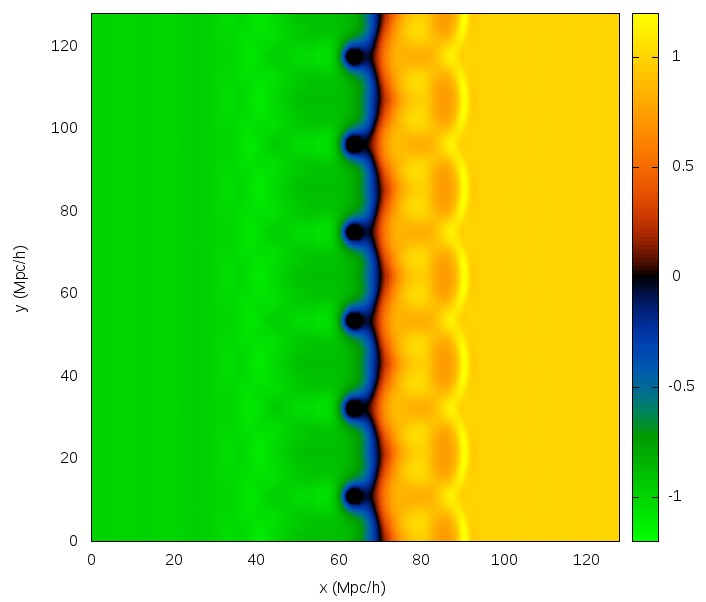}
\end{minipage}
\caption{Snapshots of a 2D simulation of the interaction of a DW with a set of filaments.  The color coding corresponds to the scalar field $\chi$.  The upper-left panel is the initial condition and time runs from top-left to bottom-right. This is an example of a wall getting trapped by the filaments.}    
\label{fig:snapshots_2d}
\end{figure*}

The first simple example consist of a planar DW in vacuum interacting with an infinite set of filaments.  The initial configuration is shown in the upper left panel of Fig.~\ref{fig:snapshots_2d} with boundary conditions being periodic in the vertical direction. The filaments are defined as disks of $1000$ times the background density. The initial configuration of the scalar field in the presence of filaments is obtained by solving the static equation of motion using a multi-grid solver \cite[e.g.][]{2014A&A...562A..78L}.  The DW is included afterwards using the analytic solution for walls in vacuum.  The initial time derivatives of the scalar field were chosen so that the wall moves to the right with the speed of light. 

To test the ability of the filaments to capture the wall and stop its movement, we made a first run with parameters $(z_{SSB}, \lambda_0)=(1, 1 \mathrm{Mpc}/h)$.  The box size used for the run was 128 Mpc/h and the grid contained 512 nodes per dimension.  The initial redshift of the simulation was $z=1$.  We used 6 equidistant filaments and, thus, the distance between them was 21.3 Mpc/h. For this particular set of parameters, the wall is not trapped by the filaments, but continues its way to the right with a reduced speed (part of the initial kinetic energy is lost into scalar waves produced during the collision).

We made further runs changing the details of this original setup and found that the wall can be trapped by the filaments by doing separately any of the following things:
\begin{enumerate}
\item Increase the number of filaments in the box from 6 to 12 ({\it i.~e.} giving them a distance of 10.6 Mpc/h).
\item Decrease the initial speed of the wall from $c$ to $0.5c$
\item Increase the radius of the filaments from 1 to 2 Mpc/h.
\item Decrease the range of the field (while adjusting $z_{SSB}$ according to the solar system constraints).  The model tested had $(z_{SSB}, \lambda_0)=(4, 0.25 \mathrm{Mpc}/h)$. For that particular run we also increased the resolution to 1024 nodes per dimension.
\end{enumerate}
We tried to trap the wall by increasing the density of the filaments, but the scalar field is already screened with $\eta=1000$, so adding more matter does not change anything in the solution.  To study the impact of the density we repeated the run with filaments of 2 Mpc/h radius, but using a density $\eta=100$ instead of 1000, and found that the wall was no longer trapped by the filaments.

Fig.~\ref{fig:snapshots_2d} shows an example of a wall that is trapped by the filaments (time runs from top-left to bottom-right).  This is the case 3 from above, with filaments of radius 2 Mpc/h.  It is possible to see the wall stretching in the regions between the filaments and then being pulled back to a configuration for which the energy is minimized.  Some of the original energy is lost in scalar waves traveling away from the wall to the right. 

In summary, a planar moving DW can be trapped by a set of filaments of constant density.  The parameters that determine the outcome are the density, the radius and the spacing of the filaments, the initial velocity of the wall and the range of the field.

\subsubsection{One filament and a bent domain wall}

\begin{figure*}
\begin{minipage}[t]{0.49\textwidth}
\includegraphics[width=1.0\textwidth]{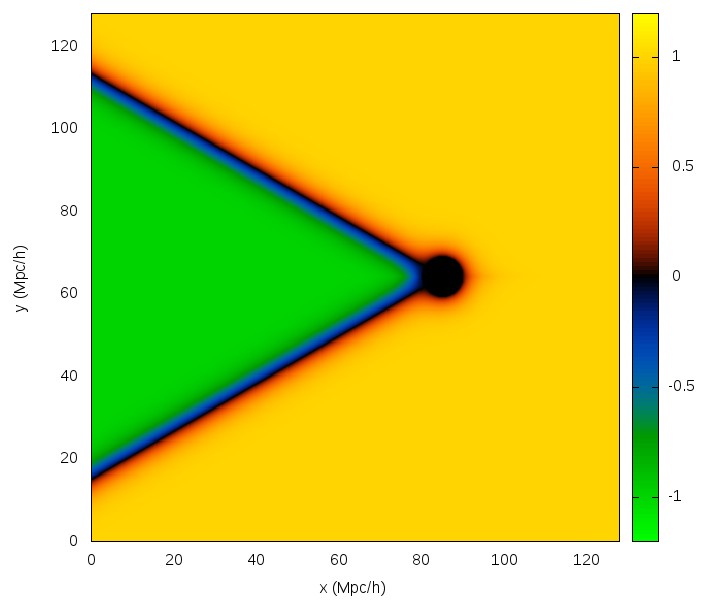}
\end{minipage}
\hfill{}
\begin{minipage}[t]{0.49\textwidth}
\includegraphics[width=1.0\textwidth]{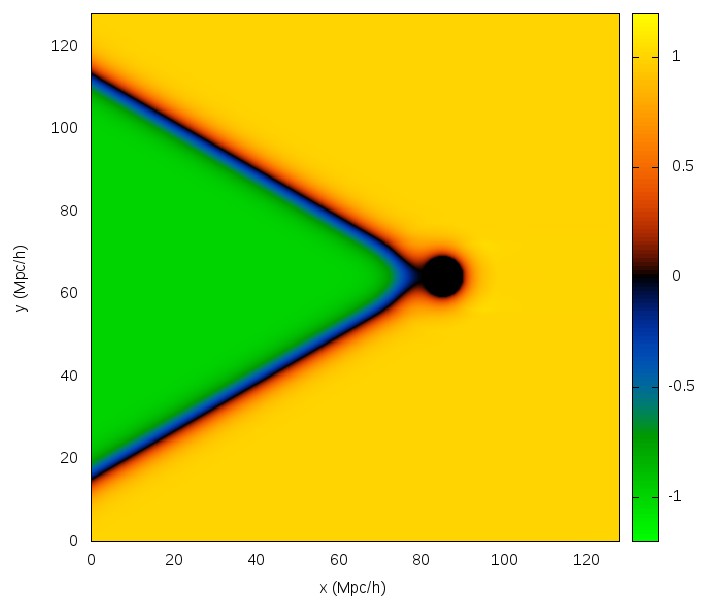}
\end{minipage}
\begin{minipage}[t]{0.49\textwidth}
\includegraphics[width=1.0\textwidth]{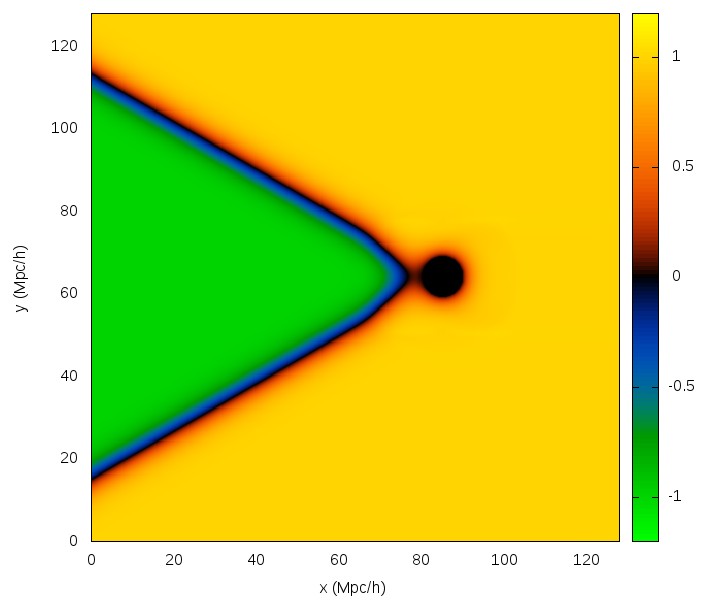}
\end{minipage}
\hfill{}
\begin{minipage}[t]{0.49\textwidth}
\includegraphics[width=1.0\textwidth]{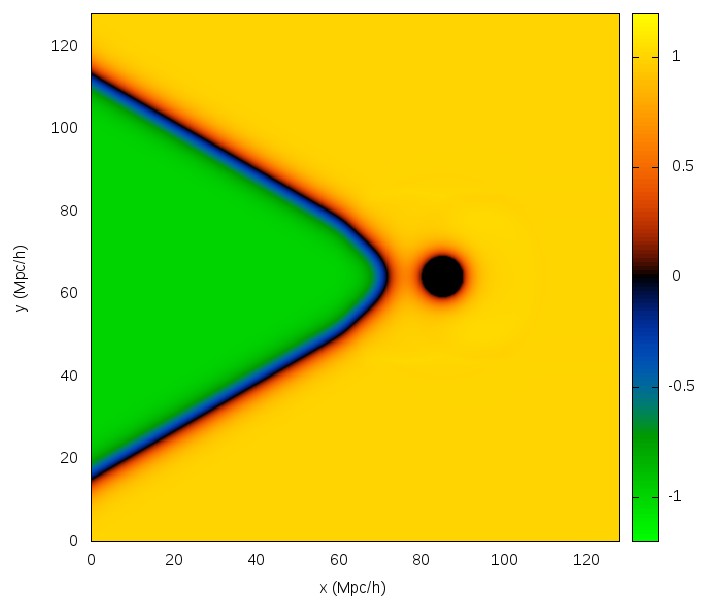}
\end{minipage}
\caption{Snapshots of a 2D simulation of unpinnned DW moving away from a filament.  The color coding corresponds to the scalar field $\chi$.  The upper-left panel is the initial condition.}    
\label{fig:snapshots_2d_two_walls}
\end{figure*}

In order to test how the pinning of the walls depends on their shape, we ran a 2D simulations using an initial configuration shown in the upper left panel of Fig.~\ref{fig:snapshots_2d_two_walls}. The system consist of one filament of radius $R$ and density $\eta=1000$, and a bended DW that passes through the filament. The angle $\alpha$ between the upper and lower parts of the wall is a free parameter.  The initial time derivative of the scalar field is zero. 

In the absence of the filament, the wall will act to decrease its surface.  As the wall is fixed to the left border of the box by the choice of boundary conditions, it will decrease its surface by pushing itself towards the left border.  The aim of the runs is to show that the filament can stabilize the original configuration of the wall. The outcome depends on the angle between the upper and lower sides of the walls.  For smaller values of $\alpha$, the wall behaves as if it was in vacuum.  The rest of the panels in Fig.~\ref{fig:snapshots_2d_two_walls} show the evolution of a wall that decouples from the filament.

Table \ref{tab:stability_two_walls} shows the results of the runs. P and U correspond to pinned and unpinned outcomes.  The runs were made with two different sets of model parameters: $(z_{SSB}, \lambda_0)=(1, 1 ~\mathrm{Mpc}/h)$ and $(10, 0.08 ~\mathrm{Mpc}/h)$.  The runs with smaller values of the range $\lambda_0$ results in pinning at smaller angles between the walls.

In summary, a bent DW can stay attached to a filament assuming that the angle between both sides of the wall is large enough.  Increasing the radius of the filament or reducing the range of the field makes the system more stable.

\begin{table}[h]
  \begin{center}
    \begin{tabular}{|c|c|c|c|c|} \hline
      \multicolumn{5}{ |c| }{$(z_{SSB}, \lambda_0)=(1, 1 ~\mathrm{Mpc}/h)$} \\
      \hline
      \backslashbox { R (Mpc/h) }{ $\alpha (^{\circ})$ } & 60 & 90 & 120 & 150 \\ \hline
      1 & U & U & U & P \\ \hline
      5 & U & P & P & P \\ \hline
    \end{tabular}
    
    \begin{tabular}{|c|c|c|c|c|c|} \hline
      \multicolumn{6}{ |c| }{$(z_{SSB}, \lambda_0)=(10, 0.08 ~\mathrm{Mpc}/h)$} \\
      \hline
      \backslashbox { R (Mpc/h) }{ $\alpha (^{\circ})$ } & 30 & 60 & 90 & 120 & 150 \\ \hline
      1 & U & P & P & P & P \\ \hline
      5 & P & P & P & P & P \\ \hline
    \end{tabular}
    \caption{Outcome of runs with a bent domain wall initially passing through a filament for different values of the angle $\alpha$ and the radius $R$ for two different sets of model parameters.  The letters P and U correspond to pinned and unpinned walls at the final time.  
}
    \label{tab:stability_two_walls}
  \end{center}
\end{table}

\subsubsection{Polygonal arrangement and resolution effects}

Next, we consider a set of filaments uniformly distributed around a circumference and a polygonal DW connecting them.  A similar configuration was studied in Ref.~\cite{2014arXiv1409.6570P}. We found that the wall becomes unpinned for low number of filaments and remains pinned after increasing their number above a certain number (more filaments imply greater angle between the sides of the polygonal wall and thus, a more stable configuration).

In this particular experiment we found that the resolution used to run the simulation plays an important role for the stability of the walls. For instance, we found that a system with four filaments of 5 Mpc/h radius which is stable at high resolutions can become unstable when the resolution is reduced below a certain value. This dependence on the resolution depends also on the size of the system (large systems tend to be more stable even at the lower resolutions that we considered). 

In light of the above findings, we caution that the domain walls seen in cosmological simulations \cite{2013PhRvL.110p1101L} could turn out to be stable if sufficient resolution was used.  Higher resolution cosmological simulations are required to test if the collapse of the wall presented in \cite{2013PhRvL.110p1101L} was a real or a resolution effect. 

\section{Conclusions}

There is a vast literature on detailed analytical and numerical studies of dynamics, interactions, evolution and gravitational effects of conventional domain walls. On the other hand, very little is known about domain walls in the case of scalar field(s) coupled to matter.  In this paper, we studied properties of domain walls in a specific model of a non-minimally coupled scalar field -- the symmetron model.

We found that the width and the surface energy of symmetron domain walls depend on the matter density and, therefore, evolve with the change in cosmological density. Furthermore, we found that the energy fraction in domain walls can be made arbitrarily large by an appropriate choice of symmetron parameters without violating the existing constraints obtained without taking the domain wall formation into account. In order to estimate the cosmological density of domain walls, we introduced a semi-analytical description of wall formation which differs from the mechanism responsible from standard walls. The average distance between domain walls is approximately of the order of horizon size at the time of the symmetry breaking.

Our results show that domain walls born well after matter-domination equality contribute a small fraction to the total energy density. However, $\Omega_{\rm DW}$ becomes large for large values of $z_{SSB}$.  This would leave an imprint on the evolution of the background, not to mention the linear and non-linear growth of structures. Thus, taking domain wall formation into account provides an upper bound to the model parameter $z_{SSB}$, for which the present literature does not provide any constraint. 

We also studied some important differences that exist in between conventional and symmetron walls which are related to their dynamics near matter overdensities. Cosmological N-body simulations that included the non-static evolution of the scalar field \cite{2013PhRvL.110p1101L, 2014PhRvD..89h4023L} showed for the first time that symmetron domain walls trace the distribution of dark matter halos.  These simulations also showed that even though domain walls followed halos and could adopt stable configurations, they eventually collapsed as time passed.  Here we studied the conditions for stability in more detail.

On the analytical side, we provide a stability criteria for spherical domain walls coupled to matter, which depends on the gradient of the density in the radial direction.  Furthermore, we studied the stability of the walls using 2D simulations in idealised conditions.  Realistic domain walls that would form in cosmological contexts are not expected to have spatial symmetries of any kind and thus, it is not easy to identify a single condition for their stability. The simulations presented in this work were limited to controlled conditions which allowed us to isolate the separate effects responsible for stability of the walls.  These include dependence on parameters of the model itself (such as the range of the field) as well as the geometry of the walls and the distribution of the matter density.  For instance, we found that bent domain walls can become unstable and decouple from overdense region when their curvature exeeds a specific value.  We also found that a travelling wall can be trapped by a set of filaments for certain spacings and radii of the filaments.

Finally, we studied the impact of resolution in the stability of the walls and found that indeed, stable domain walls can appear as unstable when the spacial resolution used during the simulations goes below a specific limit.  Thus, the question about the stability domain walls found in cosmological simulations remains open.  Further simulations and convergence tests for instance on the scaling relations for the walls that can be extracted from the simulations are required to give a definitive answer on this matter.

In conclusion, domain walls in non-minimally coupled models can have interesting novel properties not possessed by conventional walls. With recent advances in understanding the viability of scalar-tensor theories \citep{2009PhRvD..79f4036N, 1993PhRvD..48.3641B, 1974IJTP...10..363H, 2007PhRvD..76f4004H}, it is possible that there will be other examples of interesting models which allow for topological defect solutions. One particular modified gravity model that will certainly lead to topological defects is the vector-tensor model proposed in \cite{2013PhLB..725..212B}, which includes a symmetron screening mechanism for the vector field. Our study has not, by any means, exhausted the study of domain walls in the symmetron model. Future studies will include numerical simulations of wall-overdensity interactions in 3D and higher resolution cosmological simulations of structure formation in symmetron models. 

\acknowledgements CLL acknowledge support from the Research Council of Norway through grant 216756.  LP is supported by a Discovery Grant from the Natural Sciences and Engineering Research Council of Canada. CLL and LP acknowledge hospitality at DAMTP, University of Cambridge. The simulations were performed on the NOTUR Clusters \texttt{HEXAGON}, the computing facilities at the Universities of Bergen, Norway.  We thank David F. Mota for helpful discussions.

\bibliography{references}

\end{document}